# Edge currents shunt the insulating bulk in gapped graphene

M. J. Zhu[1], A. V. Kretinin[2,3], M. D. Thompson[4], D. A. Bandurin[1], S. Hu[1], G. L. Yu[1], J. Birkbeck[1,2], A. Mishchenko[1], I. J. Vera-Marun[1], K. Watanabe[5], T. Taniguchi[5], M. Polini[6], J. R. Prance[4], K. S. Novoselov[1,2], A. K. Geim[1,2*], M. Ben Shalom[1,2*]

[1]School of Physics and Astronomy, The University of Manchester, Manchester M13 9PL, UK
[2]National Graphene Institute, The University of Manchester, Booth St. E, Manchester M13 9PL, UK
[3]School of Materials, The University of Manchester, Manchester M13 9PL, UK
[4]Department of Physics, University of Lancaster, Lancaster LA1 4YW, UK
[5]National Institute for Materials Science, 1-1 Namiki, Tsukuba 305-0044, Japan
[6]Istituto Italiano di Tecnologia, Graphene labs, Via Morego 30I-16163, Italy

**An energy gap can be opened in the electronic spectrum of graphene by lifting its sublattice symmetry[1-4]. In bilayers, it is possible to open gaps as large as 0.2 eV. However, these gaps rarely lead to a highly insulating state expected for such semiconductors at low temperatures[5-11]. This long-standing puzzle is usually explained by charge inhomogeneity[6-10]. Here we investigate spatial distributions of proximity-induced superconducting currents in gapped graphene and, also, compare measurements in the Hall bar and Corbino geometries in the normal state. By gradually opening the gap in bilayer graphene, we find that the supercurrent at the charge neutrality point changes from uniform to such that it propagates along narrow stripes near graphene edges. Similar stripes are found in gapped monolayers. These observations are corroborated by using the 'edgeless' Corbino geometry in which case resistivity at the neutrality point increases exponentially with increasing the gap, as expected for an ordinary semiconductor. This is in contrast to the Hall bar geometry where resistivity measured under similar conditions saturates to values of only about a few resistance quanta. We attribute the metallic-like edge conductance to a nontrivial topology of gapped Dirac spectra[12-14].**

The gapless spectra of mono- and bi- layer graphene (MLG and BLG, respectively) are protected by symmetry of their crystal lattices. If the symmetry is broken by interaction with a substrate[3,4] or by applying an electric field[1,2], an energy gap opens in the spectrum. In BLG, its size $E_{gap}$ can be controlled by the displacement field $D$ applied between the two graphene layers. Large gaps were found using optical methods[5] and extracted from temperature ($T$) dependences of resistivity $\rho$ at sufficiently high $T$ [6-10]. Their values are in good agreement with theory. On the other hand, at low $T$ (typically, below 50 K), $\rho$ at the charge neutrality point (CNP) in gapped graphene is often found to saturate to relatively low values that are incompatible with large $E_{gap}$[6-11]. This disagreement is attributed to remnant charge inhomogeneity[6,8,10] that results in hopping conductivity and, therefore, weakens $T$ dependences. Alternative models to explain the subgap conductivity were proposed, too. They rely on the nontrivial topology of Dirac bands in gapped MLG and BLG[12-15], which gives rise to valley-polarized currents[13-15]. Large nonlocal resistances were reported for both graphene systems at the CNP and explained by valley currents propagating through the charge-neutral bulk[16-18]. Graphene edges[12,15], p-n junctions[14,19] and stacking boundaries[20] can also support topological currents. These conductive channels were suggested to shunt the insulating bulk, leading to a finite $\rho$. Experimentally, the situation is even more complicated because additional conductivity may appear for trivial reasons such as charge inhomogeneity induced by chemical or electrostatic doping[21-23]. Here we show that highly conductive channels appear near edges of charge-neutral graphene if an energy gap is opened in its spectrum. We tentatively attribute the edge channels to the presence of such unavoidable defects as, e.g., short zigzag-edge segments[12]. Their wavefunctions extend deep into the insulating bulk where they sufficiently overlap to create a quasi-one-dimensional impurity band with little intervalley scattering and high conductivity. We believe that, in certain graphene devices, the localization length can be very long, comparable to typical distances between electric contacts, which effectively results in shunting the gapped bulk.

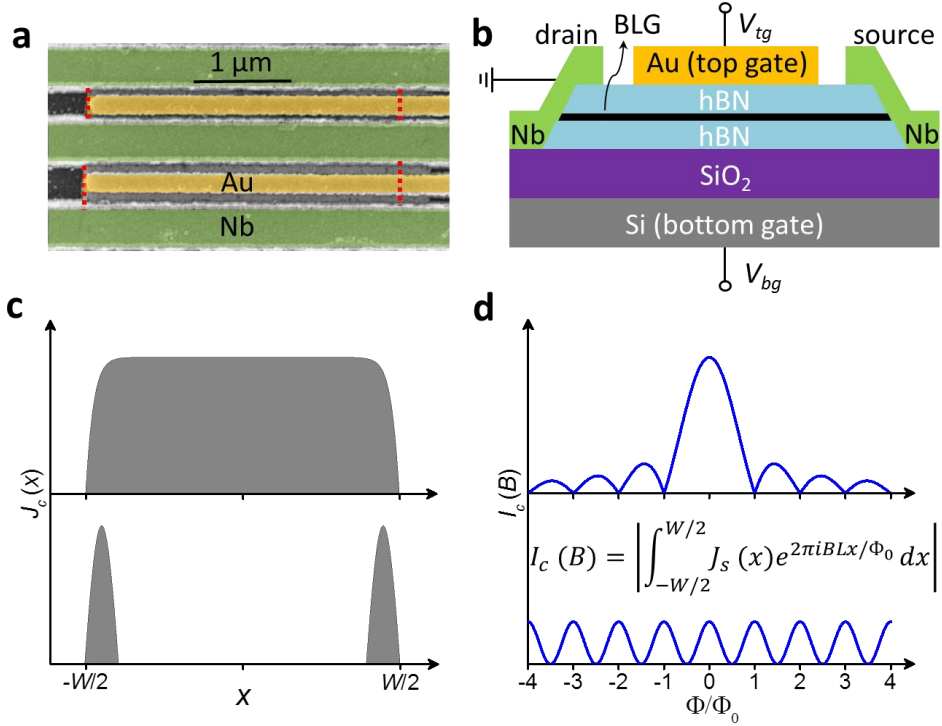

**Figure 1 | Gated Josephson junctions and spatial distribution of supercurrents. a,** Electron micrograph of our typical device (in false color). Nb leads (green) are connected to bilayer graphene (its edges are indicated by red dashes). The top gate is shown in yellow. **b,** Schematics of such junctions. **c,** Illustration of uniform and edge-dominant current flow through Josephson junctions (top and bottom panels, respectively). **d,** The corresponding behavior of the critical current $I_c$ as a function of $B$. $I_c(B)$ is related to $J_s(x)$ by the equation shown in **d**. For a uniform current flow, $I_c$ should exhibit a Fraunhofer-like pattern (top panel) such that the supercurrent goes to zero each time an integer number $N$ of magnetic flux quanta $\Phi_0$ thread through the junction. Maxima in $I_c$ between zeros also become smaller with increasing $N$. For the flow along edges (bottom panel), $I_c$ is minimal for half-integer flux values $\Phi = (N+1/2)\Phi_0$, and maxima in $I_c$ are independent of $B$. The spatial distribution $J_s(x)$ can be found[24,25] from $I_c(B)$ using the inverse FFT. Due to a finite interval of $\Phi$ over which the interference pattern is usually observed experimentally, $J_s(x)$ obtained from the FFT analysis are usually smeared over the x-axis as shown schematically in **c**.

We start with discussing behavior observed for superconductor-graphene-superconductor (SGS) Josephson junctions. Our devices were short and wide graphene crystals that connected superconducting Nb electrodes[24] (Fig. 1). Each device contained several such SGS junctions with the length $L$ varying from 300 to 500 nm and the width $W$ from 3 to 5 μm. To ensure highest possible quality[24], graphene was encapsulated between hexagonal boron nitride (hBN) crystals with the upper hBN serving as a top-gate dielectric and the Si/SiO$_2$ substrate as a bottom gate (Fig. 1b). For details of device fabrication and characterization we refer to Methods and Supplementary Information (SI). By measuring the critical current $I_c$ as a function of perpendicular magnetic field $B$, the local density $J_s(x)$ in the $x$ direction perpendicular to the super-current flow can be deduced[25], as illustrated in Fig. 1c,d. This technique is well established and was previously used to examine, for example, edge states in topological insulators[26] and wave-guided states in graphene[22]. In our report, we exploit the electrostatic control of the BLG spectrum to examine how $J_s(x)$ changes with opening the gap.

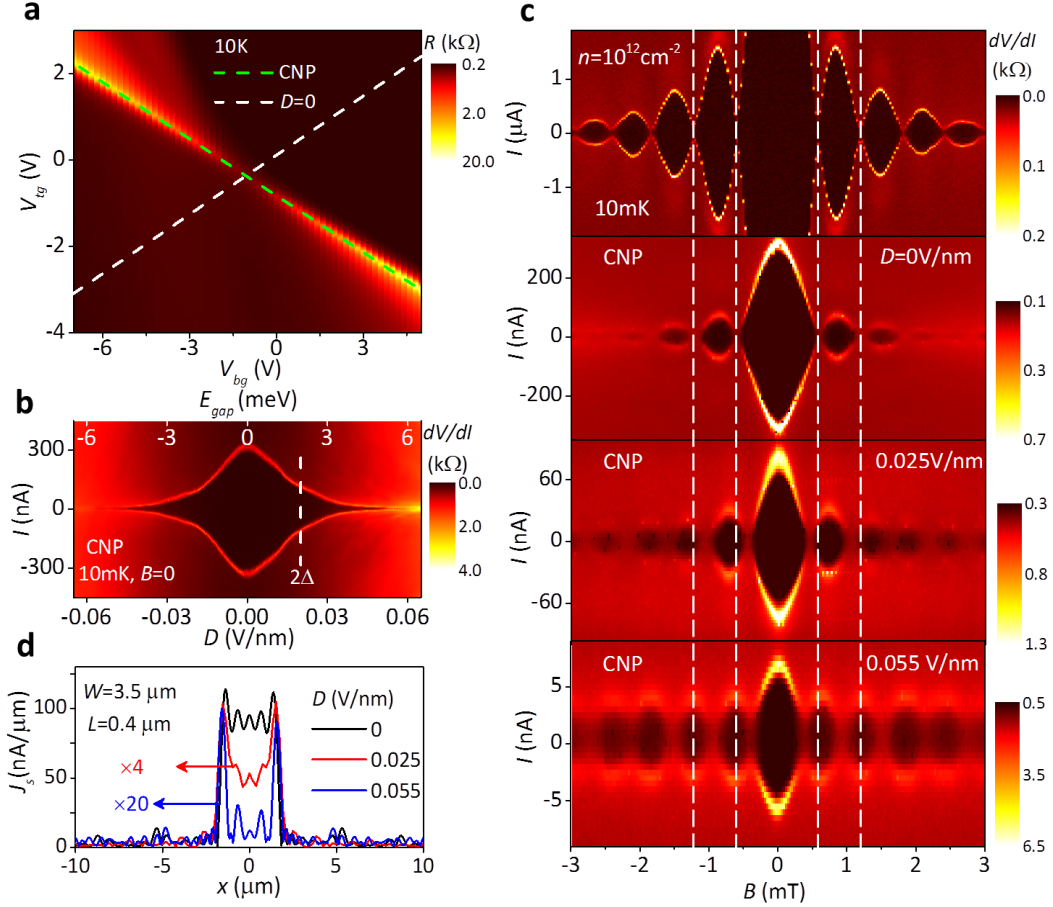

**Figure 2 | Redistribution of supercurrent as the gap opens in bilayer graphene. a,** Resistance $R$ of one of our Josephson junctions (3.5 μm wide and 0.4 μm long) above the critical $T$ as a function of top and bottom gate voltages. The dashed white line indicates equal doping of the two graphene layers with carriers of the same sign. The dashed green line marks the CNP (maximum $R$) and indicates equal doping with opposite-sign carriers. **b,** Differential resistance $dV/dI$ measured along the green line in **a** at low $T$ and in zero $B$. Transition from the dissipationless regime to a finite voltage drop shows up as a bright curve indicating $I_c$. The vertical line marks the superconducting gap of our Nb films. **c,** Interference patterns in small $B$. The top panel is for the case of high doping [$I_c(B=0) \approx 10$ μA] and indistinguishable from the standard Fraunhofer-like behavior illustrated in Fig. 1d. The patterns below correspond to progressively larger $E_{gap}$. Changes in the phase of Fraunhofer oscillations are highlighted by the vertical dashed white lines. **d,** Extracted spatial profiles of the supercurrent density at the CNP for the three values of $D$ in **c**.

By varying the top and bottom gate voltages ($V_{tg}$ and $V_{bg}$, respectively), it is possible to keep BLG charge neutral while doping the two graphene layers with carriers of the opposite sign (see Fig. 2a). This results in the displacement field $D(V_{tg},V_{bg})$ that translates directly into the spectral gap[1,2,5,6]. Its size $E_{gap}(D)$ can be deduced not only theoretically but also measured experimentally, as discussed in section 1 of SI. To quantify proximity superconductivity in our devices, we define their critical current $I_c$ as the current at which the differential resistance $dV/dI$ deviates from zero above our noise level[24]. With reference to Fig. 2b, $I_c$ corresponds to the edge of the dark area outlined by bright contours. At high doping (Fermi energy > $E_{gap}$) and low $T$, $I_c$ is found to depend weakly on $D$, reaching values of a few µA/µm, in agreement with the previous reports[22,24,27]. The supercurrent generally decreases with increasing junction's resistance and becomes small at the CNP. Its value depends on $E_{gap}$ (Fig. 2b). Accordingly, the largest $I_c$ in the neutral state is found for zero $D$ (no gap) reaching ≈300 nA for the junction shown in Fig. 2. The value drops to 2 nA at $D = \pm 0.07$ V/nm, which corresponds to $E_{gap} \approx$ 7 meV. For larger gaps, $I_c$ becomes smaller than 1 nA and could no longer be resolved because of a finite temperature (down to 10 mK) and background noise[24].

We analyze changes in the interference pattern, $I_c(B)$, with increasing $D$ (that is, increasing $E_{gap}$). At zero $D$, we observe the standard Fraunhofer pattern at the CNP, which is basically similar to that measured at high doping (cf. two top panels of Fig. 2c). Only absolute values of $I_c$ are different because of different ρ, as expected[24]. The Fraunhofer pattern corresponds to a uniform current flow (Fig. 1c,d). In contrast, the interference pattern measured at the CNP for a finite gap is qualitatively different (see Fig. 2c; $D = 0.055$ V/nm). The phase of the oscillations changes by 90° and the central lobe becomes twice narrower. In addition, the side lobes no longer decay with increasing $B$ but exhibit nearly the same amplitude. Such a pattern resembles the one shown schematically in Fig. 1d for the case of the supercurrent flowing along edges. The only difference with Fig. 1d is that in our case the central lobe remains higher than the others. For quantitative analysis, we calculated the inverse fast Fourier transform (FFT) of $I_c(B)$, which yielded[26] the current distributions $J_s(x)$ shown in Fig. 2d. The supercurrent is progressively pushed towards device edges with increasing the gap. This is already visible for $D = 0.025$ V/nm but further increase in $D$ suppresses the bulk current to practically zero, within the experimental accuracy of our FFT analysis (Fig. 2d). The accuracy is limited by a finite range of $B$ in which the interference pattern could be detected (section 2 of SI).

For completeness, we have also studied SGS junctions that were fabricated using monolayer graphene placed on top of hBN and aligned along its crystallographic axes. Such alignment (within 1-2°) results in opening of a gap of ≈30 meV at the main CNP[3,4], and secondary CNPs appear for high electron and hole doping[3,4,16]. Unlike for the case of BLG, $E_{gap}$ cannot be changed in situ in MLG devices, but one can still compare interference patterns for neutral and doped states of the same SGS junction and, also, use nonaligned junctions as a reference. Fig. 3a,b show typical behavior of $I_c$ as a function of carrier concentration $n$ for SGS devices made from gapped (aligned) and gapless (nonaligned) MLG. In the gapped device, the supercurrent is suppressed not only at the main CNP but also at secondary CNPs. For all electron and hole concentrations away from the CNPs, both devices exhibit the standard Fraunhofer pattern indicating a uniform supercurrent flow (cf. top panels of Fig. 3c,d). The same is valid at the CNP in gapless graphene (Fig. 3d,f). In contrast, for gapped MLG, the interference pattern at the main CNP undergoes significant changes such that the phase and period of oscillations in $I_c$ change (Fig. 3c; bottom panel), somewhat similar to the behavior of gapped BLG at the CNP. Quantitative analysis using FFT again shows that, in gapped MLG, the supercurrent flows predominantly along graphene edges for $n < \pm 5 \cdot 10^{10}$ cm$^{-2}$ (Fig. 3e). The figure seems to suggest a shift of conductive channels from edges into the interior. This shift originates from the increase in the Fraunhofer period at the CNP in Fig. 3c and corresponds to a decrease in the junction's effective area. However, we believe that this shift may arise from non-uniform doping along the current direction. Our MLG devices do not have a top gate and this allows doping by metal contacts to extend significantly (tens of nm) inside the graphene channel[28] which reduces the effective length of the junction.

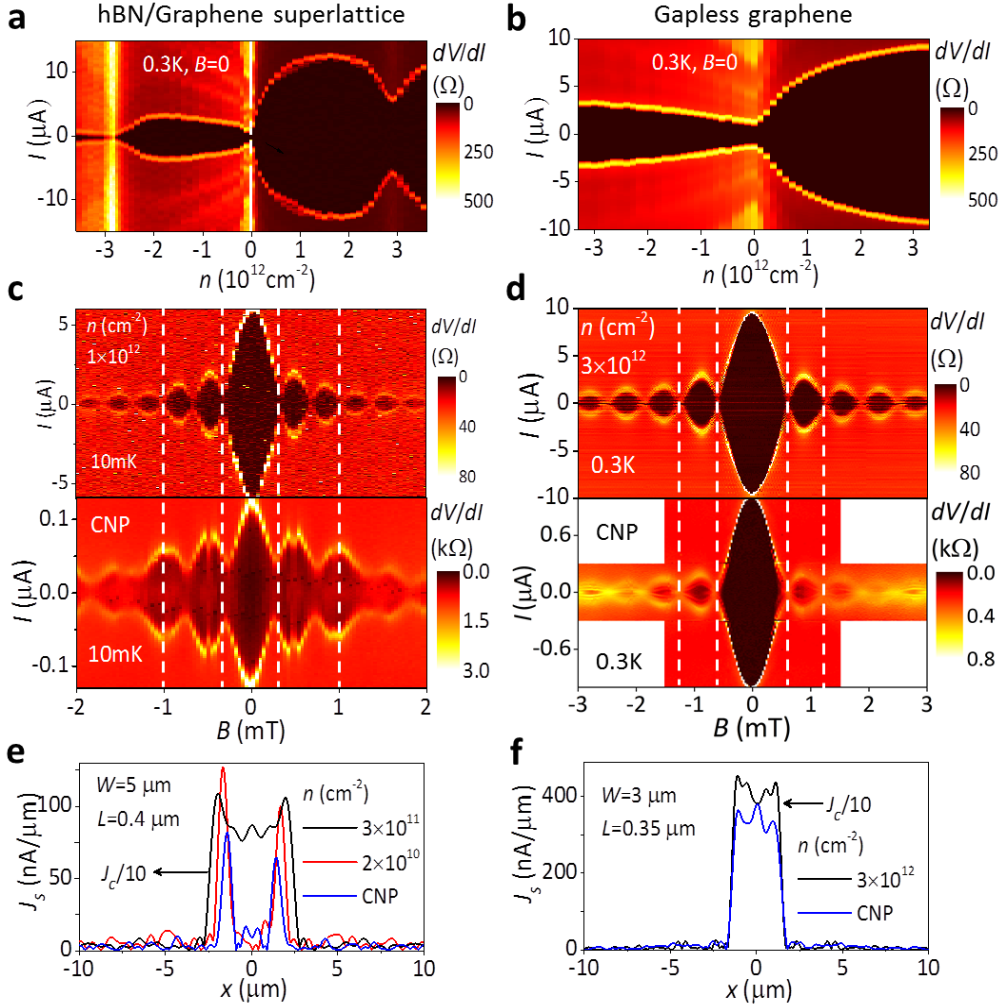

**Figure 3 | Interference patterns and supercurrent flow in gapped and non-gapped graphene monolayers. a,** Differential resistance as a function of carrier concentration $n$ and applied current $I$ for a Nb-MLG-Nb junction (5 μm wide and 0.4 μm long). The gap is induced by alignment with the bottom hBN crystal. **b,** Same for encapsulated but nonaligned monolayer graphene (the junction is 3 μm wide and 0.35 μm long). **c,** Interference patterns for gapped MLG at relatively high doping (top panel) and at the CNP. **d,** Same for non-gapped graphene. **e,f,** Corresponding spatial profiles of the current flow. They were calculated using experimental patterns such as shown in **c** and **d**. Note that graphene edges in **e** support fairly high supercurrent at the CNP whereas there is no indication of any enhanced current density along edges for the non-gapped case in **f**.

We emphasize that the observed redistribution of supercurrents towards edges is an extremely robust effect observed for all 8 gapped-graphene junctions we studied and in none without a gap (more than 10)[24]. In principle, one can imagine additional electrostatic and/or chemical doping near graphene edges[21-23], which would enhance their conductivity and, hence, favor local paths for supercurrent. This mechanism disagrees with the fact that edge supercurrents appeared independently of the CNP position as a function of gate voltage (residual doping in our devices varied from practically zero to < $10^{11}$ cm$^{-2}$) and were observed for devices with the top gate being only a few nm away from the graphene plane. The latter facilitates a uniform electric field distribution. Chemical doping at graphene edges was previously reported in non-encapsulated[21] and, also, encapsulated but not annealed devices[23]. All our devices were encapsulated and thoroughly annealed, and some of them had edges that were fully covered by top hBN rather than exposed to air (section 4 of SI). We also note our Josephson experiments yielded similar supercurrent densities at BLG edges, even in the case where the two edges were fabricated differently (one is etched as discussed above and the other cleaved and covered with hBN; see Fig 1.a). The latter observation in particular indicates little external doping along the edges. Importantly, we have found no evidence for enhanced transport along edges of similar but gapless-graphene devices. To this end, we refer, for example, to Figs 3 e,f. In the gapped MLG device, near-edge $J_s$ reaches ≈ 100nA/μm. Such supercurrents would certainly be visible in the distribution profile of the non-gapped graphene at the CNP in Fig. 3f. All the above observations point at a critical role of the presence of the gap in creating local edge currents.

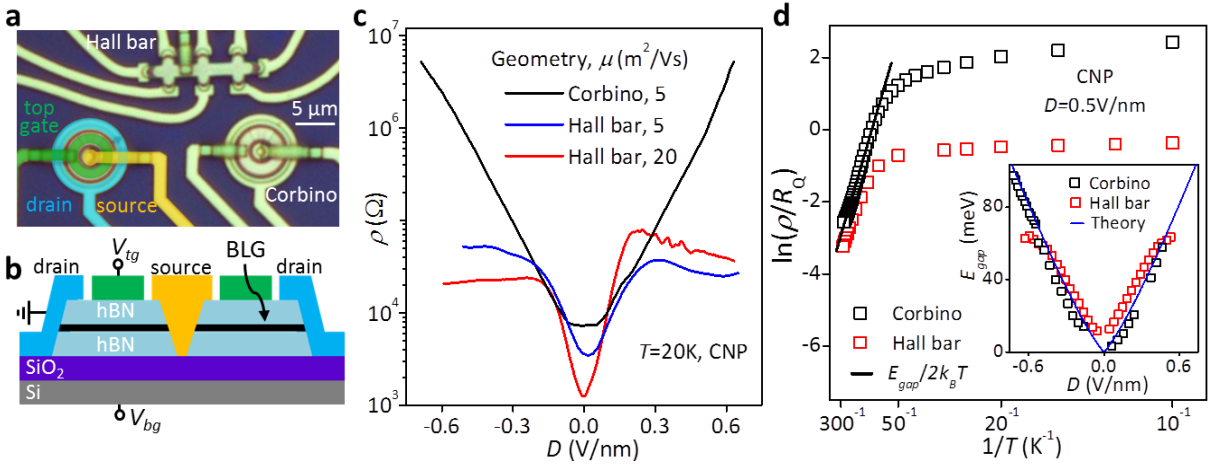

**Figure 4 | Charge-neutral bilayer graphene in the Corbino and Hall bar geometries. a,** Optical image of one of our devices with a Hall bar and two Corbino disks. The left-disk image is colored to indicate source, drain and top gate electrodes. **b,** Cross-sectional schematic of our double-gated Corbino devices. **c,** Resistivity ρ at the CNP for Corbino and Hall bar geometries as a function of $D$. For the Corbino device, ρ changes exponentially over 3 orders of magnitude. The Hall bars exhibit saturation to a few $R_Q$. **d,** Arrhenius plot for ρ($T$). The energy gap $E_{gap}$ is calculated from the linear slopes at $T > 100$ K, which are similar for both Corbino and Hall bar geometries. Below 50 K, the Hall bar device exhibits little $T$ dependence. Inset: $E_{gap}$ found for various $D$ (symbols). The blue curve is tight-binding calculations for the BLG gap from ref. 1.

While providing important insights about the current flow, Josephson interference experiments are limited to small $E_{gap}$ such that junction's resistance remains well below 1 MOhm allowing superconducting proximity. To address the situation for the larger gaps accessible in BLG devices, we compare their normal transport characteristics in the Corbino and Hall bar geometries. Because the Corbino geometry does not involve edges, such a comparison has previously been exploited to investigate the role of edge transport (for example, in the quantum Hall effect[29]). A number of dual-gated BLG devices such as shown in Fig. 4a were fabricated and examined over a wide range of $D$ and $T$. Our experiments revealed a striking difference between ρ measured in the two geometries. In the Corbino geometry, ρ at the CNP rises exponentially with $D$ and its value is limited only by a finite dielectric strength of ≈0.7 V/nm achievable for our hBN (Fig. 4b) and, at low $T$, by leakage currents. In contrast, in the Hall bar geometry, ρ at the CNP saturates at $D$ as low as < 0.2 V/nm, reaching only a few tens of kOhms at all $T$ (Fig. 4c). This disparity in the behavior of the Hall bar and Corbino devices was observed under the same measurement conditions and despite the same or higher homogeneity in the former devices. The profound difference unambiguously points at a finite conductivity caused by the presence of graphene edges, in agreement with the conclusions achieved from our Josephson experiments.

Another noteworthy distinction between the two geometries is their temperature dependences at the CNP. For $T$ above 100 K, both Corbino and Hall bar devices exhibited the same activation behavior $\rho \propto \exp(E_{gap}/2k_BT)$ as expected for a semiconductor with the gap $E_{gap}$ (Fig. 4d). Our measurements over a wide range of $D$ yielded $E_{gap}$[meV] ≈ 100×$D$[V/nm], in quantitative agreement with theory and previous reports[5] (inset of Fig. 4d). At lower $T$, resistivity of the Corbino devices continued growing and is well described by hopping conductivity that may involve both nearest-neighbor and variable range hopping[6,8-10] (Fig. 4d). On the other hand, $\rho(T)$ found using the Hall bars rapidly saturated below 100 K to values of a few resistance quantum $R_Q=h/e^2$ and changed little (by <30%) with decreasing $T$ down 2 K. The saturation behavior is similar to that observed for conductance along a p-n junction in oppositely biased BLG[19], and along walls separating BLG domains with AB and BA stacking[20].

Two possible scenarios for shunting the insulating state of gapped graphene have previously been put forward. Both rely on nontrivial topology of the gapped Dirac spectrum. One of them considers electronic states due to short zigzag segments[15] that may be present even at relatively random edges[12]. Although these states decay exponentially into the gapped bulk, their penetration length $\xi$ is very long with respect to the lattice constant $a$. For MLG and BLG, $\xi$ can be estimated as $\approx \hbar v/E_{gap}$ and $\hbar/\sqrt{mE_{gap}}$, respectively, where $\hbar$ is the reduced Planck constant, $v$ the Fermi velocity in MLG and $m$ the effective mass in BLG. For our typical gaps, $\xi$ is about 10–20 nm, much larger than $a$. This suggests that wavefunctions of isolated zigzag states should strongly overlap inside the bulk creating a quasi-one-dimensional (1D) band. Moreover, because $\xi/a \gg 1$, the wavefunctions mostly reside in the bulk where there are little defects, which ensures that impurity bands are effectively protected against backscattering. The situation resembles the modulation doping used to achieve high carrier mobilities in semiconductor quantum wells. The observed saturation of $\rho$ to $\sim R_Q$ and the long-range nonlocal resistance reported previously[16-18] imply that the mean free path along the quasi-1D channels can reach a micrometer scale for high-mobility graphene. Although numerical simulations[12] yielded zero-$T$ localization lengths at least an order of magnitude shorter than this scale, localization in the edge channels may be suppressed by a finite $T$ and electron-electron interactions that are prominent especially in low-dimensional conductors. Such delocalization effects have so far not been investigated theoretically. The invoked edge channels would be consistent with our experimental observations. Obviously, the mean free path can vary from sample to sample and strongly depend on fabrication procedures, which may explain only-weakly-saturating behavior that was reported in some gapped graphene devices[16,17,19]. In addition, there is a complementary scenario that also relies on the nontrivial topology of the gapped Dirac spectra but may not require zigzag segments. The valley Hall effect is inherent to gapped graphene and generates valley currents that flow perpendicular to applied electric field[13,16]. If injected from electric contacts into the gapped bulk, they are expected to become squeezed towards weakly-conductive edges, similar to what is known for the case of the quantum Hall effect[30] and in agreement with recent simulations for gapped MLG[14]. Lastly, let us mention another relevant suggestion that a weak confining potential at graphene edges may guide electronic states over large distances, independently of its strength[22,31]. In the latter scenario, an enhanced edge conductance is expected irrespectively of the gap size, which seems to contradict our experimental observation that there is little enhancement of near-edge supercurrent in the absence of the gap.

To conclude, our results show that the insulating state of gapped graphene is electrically shorted by narrow edge channels exhibiting high conductivity. This can explain low apparent resistivity often observed for charge-neural gapped graphene at low temperatures, especially in devices made from high quality graphene in which the bulk is expected to contribute little to either hopping conductivity or backscattering of edge modes[5-11,19]. Further experiments and theory are needed to distinguish between the possible scenarios described above and elucidate the nature of the reported edge conductance.

## Methods

**Device fabrication.** Mono- or bi-layer graphene crystals were encapsulated between hBN crystals (typically, ≈30 nm thick) using the dry transfer technique as detailed previously[32]. The hBN-graphene-hBN stack was assembled on top of an oxidized Si wafer (300 or 90 nm of $SiO_2$) and then annealed at 300 °C in a forming gas (Ar-$H_2$ mixture) for 3 hours. As the next step we used the standard electron-beam lithography to create a PMMA mask that defined contact regions. Reactive ion etching (Oxford Plasma Lab 100) was employed to make trenches in the heterostructure through the mask. We used a mixture of $CHF_3$ and $O_2$, which provided easy lift-off of PMMA, so that metal contacts could be deposited directly after plasma etching. This also allowed us to minimize contamination of the exposed graphene edges[24]. After this, for BLG devices, another metal film (typically, Au/Cr) was deposited on top of the heterostructure to serve as the top gate. In order to avoid the edges of graphene extending out of the metal gate, the latter is used as a part of the final etch-mask; the uncovered graphene between the contacts and the gate is protected by a second PMMA mask, allowing the metal gate to extend slightly at the crucial edge locations. For the Hall bar geometry, we often used an additional hBN crystal to cover the hBN-graphene-hBN stack after plasma etching, which allowed the metal film for the top gate to go over exposed graphene edges without touching them. To provide the central contact in Corbino devices, we used air bridges[33]. In some of our Josephson devices, graphene was not etched but made directly from cleaved crystals selected to have a strip-like shape. In this case, graphene edges were not exposed but fully encapsulated in hBN. Similar transport and Josephson behavior was found in all cases, independent of the variations in fabrication procedures.

**Transport experiments.** All electrical measurements were carried out in a He3 cryostat (Oxford Instruments) for *T* down to 0.3 K and, for lower *T*, in a dilution refrigerator with the base temperature of 10 mK (BlueFors Cryogenics). The differential resistance was measured in a quasi-four-terminal configuration (two superconducting leads for driving the current and the other two for measuring voltage) using a low-frequency lock-in technique. All electrical connections to our devices passed through a cold RC filter (Aivon Therma) placed close to the sample and additional AC filters were used outside the cryostats. At large displacement fields, our Corbino devices exhibited high resistivity such that the lock-in technique became inappropriate. In this case, we used dc measurements.

## Supplementary Information

### 1. Characterization of double gated bi-layer graphene.

By implementing top and bottom gate electrodes in the studied devices, it is possible control the charge carrier density $n$ and the electric displacement field $D$ between the two layers independently[1]. Below, the calibration procedure of $n(V_{tg}, V_{bg})$ and $D(V_{tg}, V_{bg})$ ($V_{tg}$ and $V_{bg}$ are the top and bottom gate voltages respectively). Examples of such measurements for Hall-bar, Josephson junction, and Corbino device geometries are presented in Fig. S1.

At first, the resistance $R$ is plotted as a function of the two gates (Fig. S1a,c,d). The sharp peak in $R$ determines the position of the charge neutrality point (CNP), Fig. S1b. The axis parallel to the charge neutrality line is determined (see black arrows, Fig. S1a), and its slope: $\Delta V_{bg}/\Delta V_{tg} \approx 2$ is equal to the capacitance ratio of the two gates $C_{tg}/C_{bg}$. Smaller $C_{bg}$ is expected for the thicker $SiO_2$ dielectric at the bottom, and requires separate characterization for each device due to the different thickness of hBN which we place on top of $SiO_2$. The negative slope ($\Delta V_{bg}/\Delta V_{tg} \approx -2$, marked by white line on Fig. S1a), corresponds to adding the same charge to both layers and changing the total $n$ while keeping a fixed $D$. In order to accurately measure $n$ we analyze the quantum oscillations in $R$ at high magnetic fields and away from CNP. From this, the capacitance (per unit of area) for each gate is extracted using: $ne = C_{tg}\Delta V_{tg} + C_{bg}\Delta V_{bg}$, and the displacement field is calculated to be: $D = (C_{tg}\Delta V_{tg} - C_{bg}\Delta V_{bg})/2\varepsilon_0$.

The energy gap $E_{gap}$ is measured independently from the Arrhenius-like activation of $R$ at high temperatures as shown in the main text, Fig. 4d. When measured at different $D$, we find $E_{gap}[\text{meV}] \approx 100 \times D[\text{V/nm}]$ to hold for all our BLG devices (see inset to Fig. 4d), in agreement with previous reports[2] and calculations[3].

The devices presented here also show saturation of the sub-gap $R$ with increasing $D$ in the Hall bar and Josephson geometry, and exponentially increasing $R$ in the Corbino geometry (Fig. S1b,d). For the latter, an additional increase in $R$ is observed at a fixed value of the bottom gate $V_{bg} \approx -3$ V (Fig. S1c). It corresponds to the CNP of the BLG at the locations in the device not covered by the top gate (see image in Fig. 4a, main text). This spatial separation of the top gate from the metal-graphene interface guarantees a negligibly-low contact resistance at high $D$ for the two-probe measurement in this geometry.

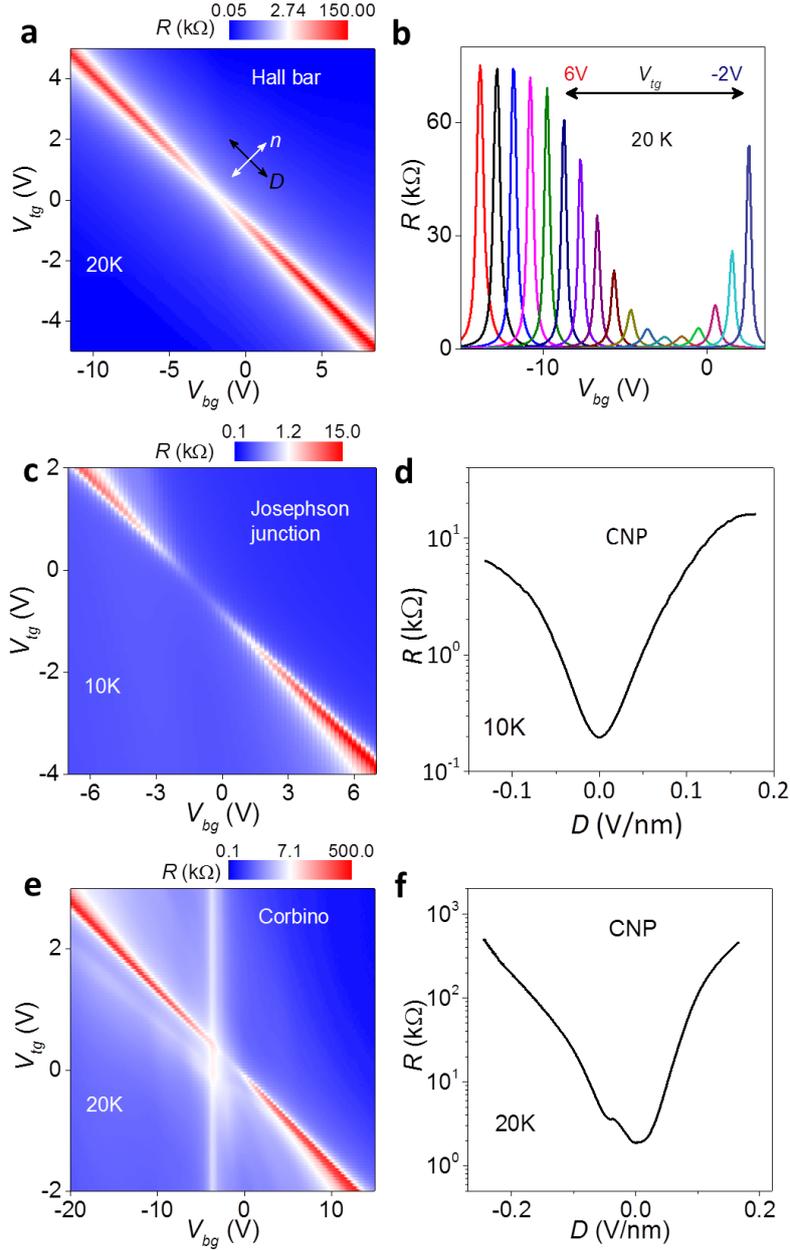

*Figure S1| Characterization of double-gated bilayer graphene. a,* Color-plots of the resistance R (in log-scale) as a function of the top and bottom gate voltages, for the Hall-bar geometry (the measured section is 2.3 μm wide and 6.6 μm long). *b,* Bottom gate scans from the map in (a) at different fixed values of the top gate. At the charge neutrality point (CNP) R is saturated for $V_{tg}$>5V corresponding to D ≈-0.2V/nm as shown in the main text (Fig. 4c). *c,* Color-plots of the resistance R (in log-scale) as a function of the top and bottom gate voltages for the Josephson junction studied in the main text Fig. 2. *d,* Resistance at CNP extracted from the map in (c). The increase in R is saturated for displacement field D≈0.15V/nm. *e,* Color-plots of the resistance R (in log-scale) as a function of the top and bottom gate voltages for a Corbino "edge-less" device. Here the top gate is 10 μm wide and 1 μm long, and it is separated by 1 μm from the inner and outer contacts. The vertical white line at $V_{bg}$≈-3V corresponds to the charge neutrality point in the part of the device which is not covered by the top gate. *f,* Resistance at CNP extracted from the map in (c). The increase in R is exponential with the displacement field D.

## 2. Example of additional BLG Josephson junction.

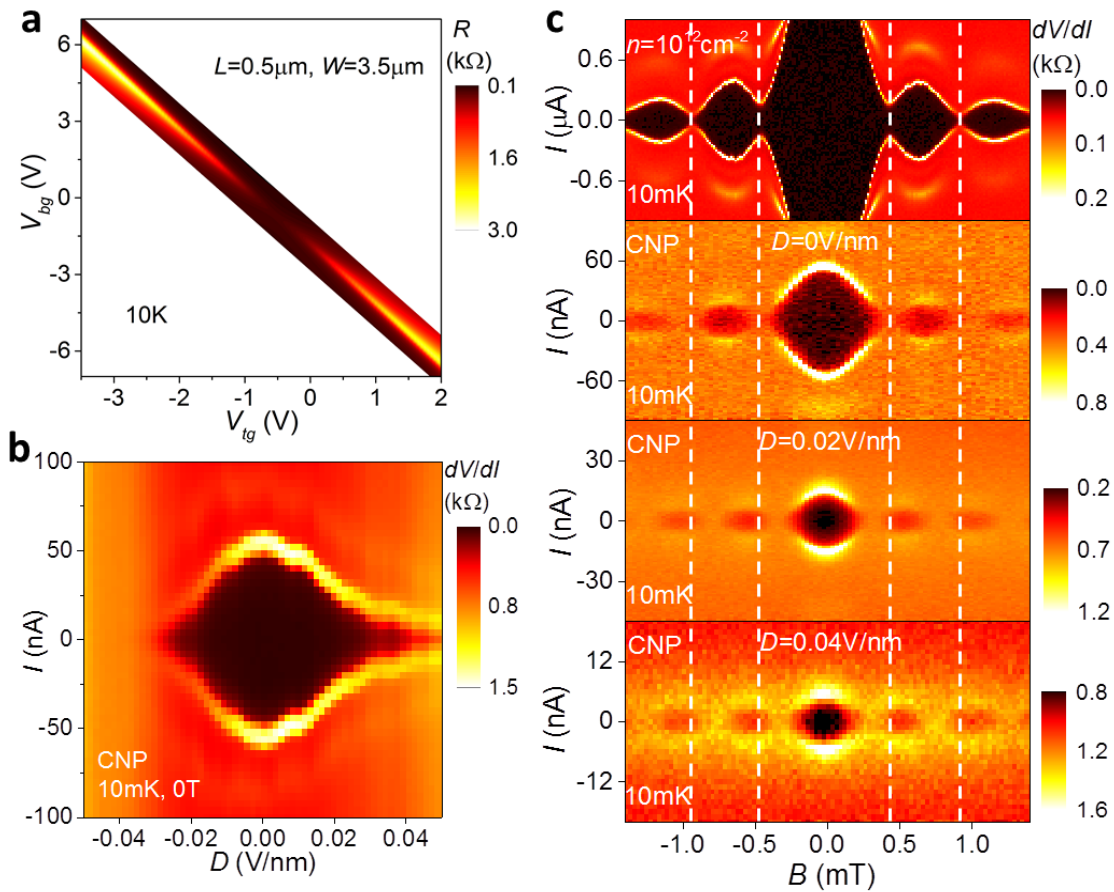

*Figure S2| Redistribution of supercurrent as the gap opens in bilayer graphene. **a**, Resistance R of a Josephson junctions (3.5 µm wide and 0.5 µm long) above the critical T as a function of top and bottom gate voltages. **b**, Differential resistance dV/dI measured along the CNP line in **a** at low T and in zero B. Transition from the dissipationless regime to a finite voltage drop shows up as a bright curve indicating $I_c$. **c**, Interference patterns in small B. The top panel is for the case of high doping [$I_c(B=0) ≈ 2$ µA] and indistinguishable from the standard Fraunhofer-like behavior illustrated in Fig. 1d. The patterns below correspond to progressively larger $E_{gap}$. Changes in the phase of Fraunhofer oscillations, consistent with the formation of edge modes, are highlighted by the vertical dashed white lines.*

## 3. Bulk versus edge distribution of the supercurrent in bi-layer graphene.

In this section we further analyze the interference patterns of the supercurrent $I_c(B)$ at CNP for different values of $D$. The inverse FFT is calculated to extract the local current distribution $J_s(x)$ (see Fig. S3b and Fig. 2d of the main text). Then the current density at the edges is compared to the one at the center of the junction. We find that the transition from uniform current distribution to the edge dominant flow is rather sharp and occupies the range in the displacement $0.015<D<0.03$ V/nm (see Fig. S3b). The bulk component of $J_s$ is dramatically reduced above $D≈0.03$ V/nm and the supercurrent becomes restricted to the edge channels. To this end we note that the supercurrent in the graphene is carried by Andreev pairs coupled by the superconductor gap Δ. At zero temperature and for entirely gaped graphene, finite $I_c$ is not expected for $E_{gap} > 2Δ$ because tunneling processes are improbable across this 400nm long barrier (the length of the graphene channel). The analysis of $J_s(x,D)$ below suggest that the cut-off for the bulk contribution is indeed happening at $E_{gap} ≈ 2Δ$ (=2meV in the case of these Nb contacts[4]). Thus the finite $I_c$ at the edges persisting to large gaps indicates that the edges are less gapped than the bulk, or not gapped at all.

The inverse FFT shown in Fig. S3b and Fig. 2d can be fitted by Gaussians in order to estimate the width $w$ to which the edge mode extend into the bulk (taken as the width of the peak at half maximum). Yet a limit on the spatial resolution of $J_s(x)$ arise, which can be defined by the largest number of the magnetic flux in which the interference pattern $I_c(B)$ still can be detected (additional limitation of the calculation is the assumption of a sinusoidal current-phase relation, which is not accurate in these long and ballistic Josephson junctions). We can reliably extract the interference over ≈10 periods (flux quanta) before the noise level or other ballistic effects[4] alters its pattern. This number correspond to a spatial resolution limit of ≈W/10=350 nm for the studied junctions of the width $W$. The calculated $w$ from the FFT is 650nm and 450nm at $D$=0.025 V/nm and 0.055 V/nm respectively and should be regarded as an upper limit of the width of the edge channels.

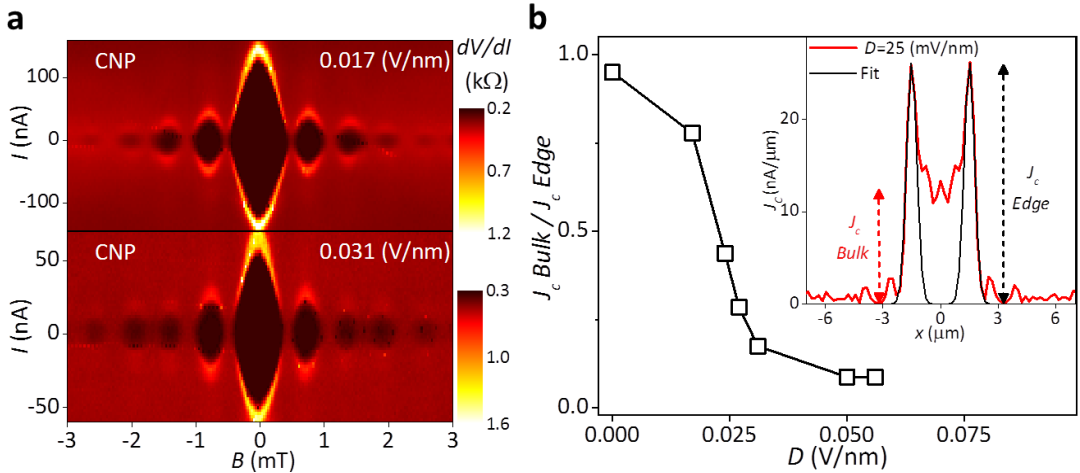

*Figure S3│ **Supercurrent distribution as a function of the displacement field D**. **a,** Examples of interference patterns measured at two different D. **b,** The supercurrent density at the bulk and at the edges is extracted from inverse FFTs of the $I_c(B)$ patterns (as shown in Fig. 2d). The transition from bulk dominant to edge current is sharp, in the range D≈0.015 to 0.03 V/nm. Inset, Gaussian fit to the edge current distribution.*

## 4. Chemical and electrostatic doping at the edge.

In principle, external doping near graphene edges may offer an alternative explanation for edge-transport when the bulk is gaped. In the following we consider various doping scenarios, how to minimize their effect and how to test its presence experimentally. The three doping scenarios are: i) Chemical variations at the edge, which may depend significantly on the fabrication process[5]. To minimize its effect we anneal all samples as an essential part of our fabrication procedures. ii) Electrostatic doping arising from a finite separation between the gate electrodes and graphene[6]. The spatial range of this stray doping is determined by the distance to the gates, which for this reason were fabricated as close as possible to the graphene plane (≈30 and 120nm away for the top and bottom gates, respectively). iii) Non-uniform termination of the two layers in the BLG. This is avoided by dry etching the two layers simultaneously using a highly anisotropic etching process.

To evaluate the effect of external doping, we measured devices in which the two edges of the Josephson junction were different (see Fig. 1a, Fig. S4). One edge of the BLG is encapsulated by hBN and overlaid by the top gate, while at the other edge the top gate terminates and the BLG edge is uncovered. In principle, the different profiles should result in different chemical and electrostatic doping. Calculations of the electrostatic doping profiles are shown for the two edge configurations (Fig. S4). When the top gate terminates above the graphene edge, the charge density accumulation is diverging near it, with a lateral cutoff given by the thickness of the dielectric spacer. 100 nm away from the edge, the carrier density is expected to be ≈$5 \times 10^9$ cm$^{-2}$ for 1V / 0.24V applied to the bottom / top gate, respectively (corresponding to $D≈0.03$V/nm). In contrast, the configuration of extended gate and hBN show negligible electrostatic doping. The effect of electrostatic or chemical doping, if significant, should clearly favor edge conductance along one of the edges only.

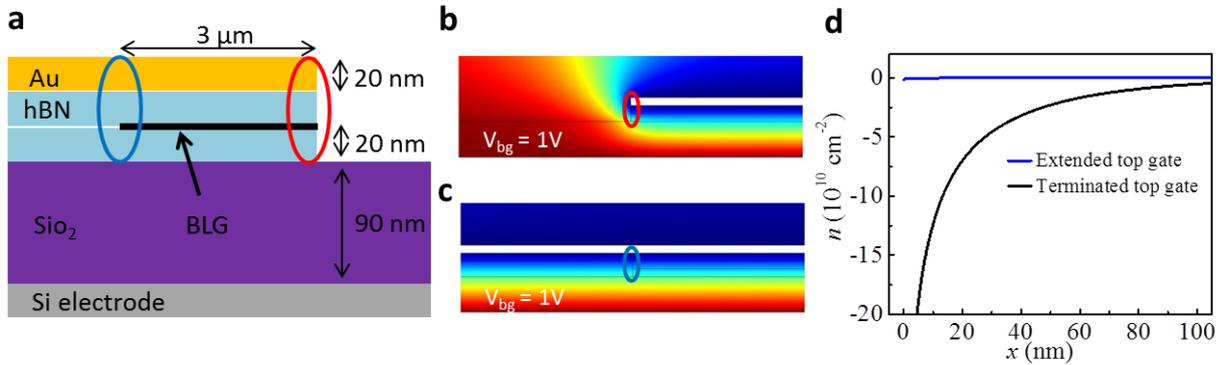

*Figure S4* | *Electrostatic modeling of edge doping. **a**,* Schematic cross-section of a Josephson junction with different edge profiles (the other cross-section and top view are shown in figure 1 a,b main text). ***b, c,*** Finite element calculation of the electrostatic potential distribution for the two edge configurations. The bottom gate is fixed at 1 Volt while the top gate is tuned to fix a zero potential at the bulk of the BLG (colored circles mark the two edge configurations) ***d,*** Calculated carrier density accumulation as function of the distance from the edge for the extended (Blue curve) and edge-terminated (red) profiles*.* For the former, charge accumulation is negligible.

Here we point out the high sensitivity of the supercurrent interference patterns to asymmetric supercurrent density distribution. Conceptually this sensitivity can be described as follow: if the maximum supercurrent density in the two edges is precisely equal, flux penetration can force it in opposite directions for each edge, such that a zero net supercurrent can be driven across the junction (the measured $I_c$). On the other hand, uneven critical current density will preserve a finite

"net" supercurrent in the better conducting edge, even when the flux-driven supercurrent at the less conducting edge is maximal. This will result in a non-zero net supercurrent flowing across the junction, at all values of magnetic flux. In the interference pattern, it will show up as a lifting of the minimum $I_c$ [7,8]. Furthermore, in the case of supercurrent flowing only in one of the edges the period of the oscillations will increase significantly, reflecting the confined width of the supercurrent and the small effective area of flux penetration.

The fact that the interference of the BLG junctions drops to zero at half integer values of flux (see Fig. 2c, Fig. S2c), indicates that the conductance at the two edges is very similar, and that the gate electrode profile does not have a significant effect on the edge modes observed.

To test the electrostatic doping scenario in the Hall bar devices, we compared top gates terminated at the edge of bilayer graphene (see Fig. 4a, main text), or extend far beyond the bilayer (see Fig. S5a). For both types of devices the sub-gap resistance at high $D$ was measured and similar saturation of $R$ was observed (Fig. S5b). It indicates again that the edge profile and the resulting external doping is not significant in these devices.

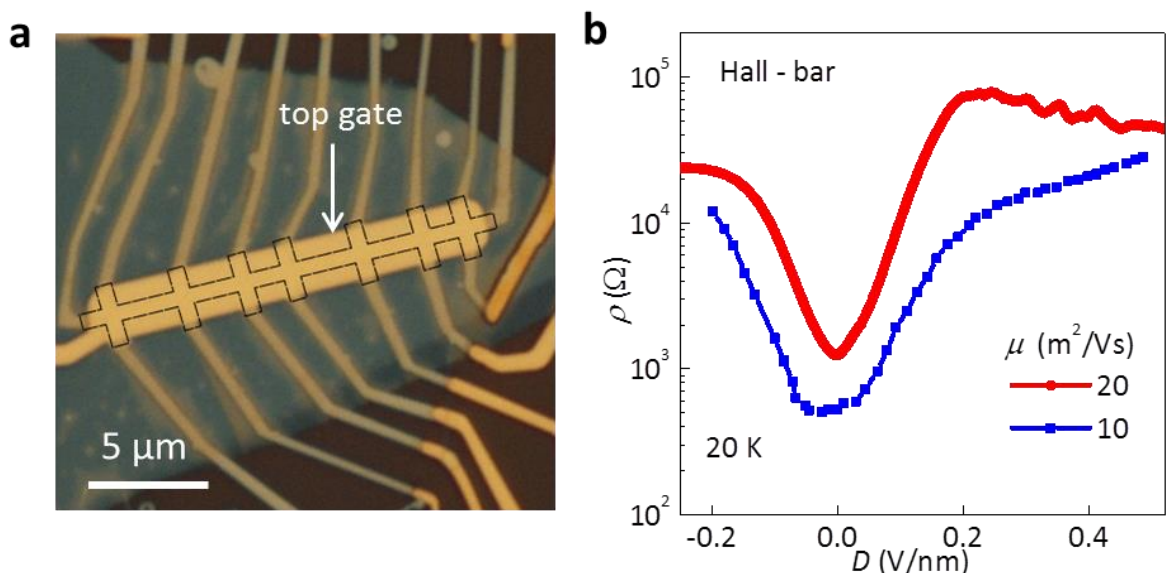

*Figure S5 | Sub-gap resistivity of BLG with the top gate extending above the edges. **a,** Optical image of the device. Additional hBN cover-layer was placed, enabling the extension of the top gate away from the BLG edges (marked by dashed black line). **b,** Resistivity ρ as a function of the displacement field measured at neutrality point for the device shown in **a** (blue curve). The exponential increase in resistivity is dumped above D≈0.2 V/nm, where ρ becomes comparable to the quantum of resistance. Devices with gate electrode terminated at the edge (red curve) show a more pronounced saturation, presumably owing to the higher mobility achieved.*

We also point to the experiments on the gapped monolayer graphene discussed in the main text. There we compare Josephson junctions made using the same fabrication procedures and geometries (including the thickness of the dielectric materials) but for non-aligned (non-gapped) and aligned (gapped) devices. Any inhomogeneity in the external doping should be essentially the same for the two cases. After testing more than 10 non-gapped monolayers[4] and 4 gaped (hBN-aligned) junctions we note that no edge current enhancement was observed in any of the former, while clear edge-dominant currents where observed in all the latter. Here, to avoid the case of the edge-modes being masked by the bulk currents, we examined different aspect-ratios of un-gapped junction with different normal state resistance above the Nb transition temperature or at currents above $I_c$ (see Fig. 3d,f). We note that no sign of edge currents was found even when the normal state resistance exceeds the resistance where edge-dominant transport was observed in the gapped graphene. It points again to the crucial role of the gap in supporting the enhanced edge conductivity rather than external doping mechanism.

## 5. On-off ratio in gapped bi-layer graphene.

Achieving high on-off ratio in gapped graphene devices is a focus of intense research driven by the practical requirements of electronic applications like field effect transistors (FET)[9]. Owing to the ballistic transport over micron length scales in pristine graphene and BLG at room temperature, the "on" state resistance is mostly determined by the metal-graphene interface resistance, which can be as low as 35 Ohm×µm[4].

The "off" state resistance is usually determined by the size of the gap and the device inhomogeneities. As has been discussed in the main text, for sufficiently clean bilayer graphene devices the edge conductance limits the sub-gap $R$ to the order of the quantum resistance. In the Josephson junction FET geometry for example, the on-off ratio is limited to $10^2$ at $D$=0.2V/nm and saturates for higher displacement fields. In contrast, for the edgeless Corbino geometry the highly resistive "off" state is recovered. Here we demonstrate on-off ratio ≈$10^4$ (at 20K), achieved already at $D$=0.2V/nm owing to the high device homogeneity. Importantly the "off" resistance is limited only by the device quality and the achievable $D$.

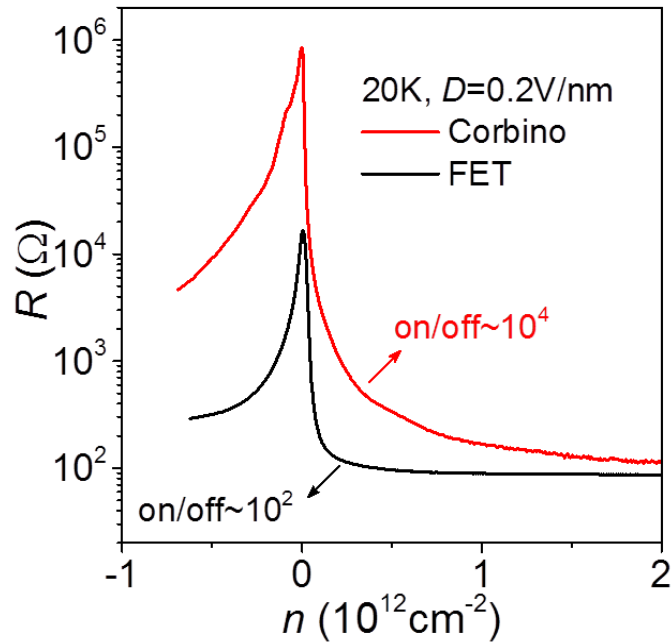

*Figure S6│ On-off ratios in Cobino and in the field effect transistor (FET) geometry. The resistance of the device in the FET geometry (3.5µm wide, 0.4µm long) changes by only 2 orders of magnitude, due to the edge-conductance at the charge neutrality point. In the case of the "edge-less" Corbino geometry, R changes by over 4 orders of magnitude already at D=0.2V/nm.*